\documentclass[11pt]{article}
\usepackage{amsmath,amssymb,epsfig,cite}
\usepackage {color}
\usepackage{bbold}
\advance\hoffset by -1.1truecm
\advance\voffset by -1.0truecm
\newcommand{\be}{\begin{equation}}
\newcommand{\ee}{\end{equation}}
\newcommand{\bea}{\begin{eqnarray}}
\newcommand{\eea}{\end{eqnarray}}

\numberwithin{equation}{section}

\newcounter{appendice}

\begin{document}

 \title{\vspace{-3cm}
 \begin{flushright}
 \end{flushright}
\vspace{2.2cm}\begin{flushleft} \bf {Monopoles On $S^2_F$ From The Fuzzy Conifold
\linethickness{.05cm}\line(1,0){433}
}\end{flushleft}}
\author{\bf{Nirmalendu Acharyya\footnote{nirmalendu@cts.iisc.ernet.in},\, and Sachindeo Vaidya\footnote{vaidya@cts.iisc.ernet.in}} \\ 
\begin{small}{\it Centre for High Energy Physics,  Indian Institute of Science, Bangalore 560012, India}
\end{small} \\
}
\date{\empty}

\maketitle

\abstract{The intersection of the conifold $z_1^2+z_2^2+z_3^2 =0$ and $S^5$ is a compact 3--dimensional manifold $X^3$. We  review the  description of $X^3$ as a principal $U(1)$ bundle over $S^2$ and construct the associated monopole line bundles. These monopoles can have only even integers as their charge.  We also show the Kaluza--Klein reduction of $X^3$ to $S^2$ provides an easy construction of these monopoles. Using the analogue of the Jordan-Schwinger map, our techniques are readily adapted to give the fuzzy version of the fibration $X^3 \rightarrow S^2$ and the associated line bundles. This is an alternative new realization of the fuzzy sphere $S^2_F$ and monopoles on it.}
 \tableofcontents

\section{Introduction}
A conifold is a complex manifold with a conical singularity at isolated point(s).  In the neighbourhood of such a singular point the conifold is described by a quadratic in $\mathbb{C}^n$ \cite{Candelas:1989js}
\begin{equation}
\sum_{\alpha=1}^n z_\alpha^2 =0,
\end{equation}
where the singular point is chosen to be at the origin of $\mathbb{C}^n$. The conifold is a $(2n-2)$ dimensional space $Y^{2n-2}$ which is smooth everywhere except at $z_\alpha=0$.

Here we will specialise to the $n=3$ case and show that the base of $Y^4$ is a fibre bundle over $S^2$. Interestingly our construction allows a natural fuzzification. The fuzzy version is a Jordan--Schwinger -- like map and is a new construction of the fuzzy two--sphere $S^2_F$.

Let us briefly recall the properties of $Y^{2n-2}$.
It is a noncompact Calabi-Yau manifold, with
$z_\alpha$ transforming as vectors of $SO(n)$. It  also admits an additional U(1) symmetry  $z_\alpha \rightarrow e^{i \lambda}z_\alpha$, so the symmetry group is $SO(n) \times U(1)$.
The base of  $Y^{2n-2}$  is a $(2n-3)$ dimensional manifold $X^{2n-3}$ which is the intersection of  $Y^{2n-2}$ with the $S^{2n-1}\equiv\{\bar{z}_\alpha z_\alpha=$fixed$\}$.  $Y^{2n-2}$ is a cone over $X^{2n-3}$. The base $X^{2n-3}$ is a compact Einstein manifold: $\left(R_{ij}\right)_{X^{2n-3}} = (2n-4)\left(g_{ij}\right)_{X^{2n-3}}$ while $Y^{2n-2}$ is Ricci flat \cite{Klebanov:1998hh}. 

The $ n=4 $ case has been studied in detail \cite{Candelas:1989js}. The symmetry group in this case is $SO(4)\times U(1) \simeq SU(2)\times SU(2) \times U(1)$.
  It can be  shown that $X^5$ (also sometimes called $T^{1,1})$ is a $U(1)$ fibre bundle over $S^2 \times S^2$.
 What makes this manifold interesting is its deep connection to  gauge-gravity duality (for example see \cite{{Klebanov:1998hh},{Pandozayas:2000sq}}).  The fuzzy version of the $n=4$ will be dealt seperately in another article\cite{acharyya_vaidya_ms}.

Our interest  in this article is in the $n=3$ case. The  symmetry group now is $SO(3)\times U(1)$ and the base $X^3$ is a 3--dimensional manifold \cite{Parthasarathy:2001ss}.
We will show that there exist a ``Hopf-like" map using the spin--1 representation of   $SO(3)$. This map from $X^3$ to $S^2$ explicitly brings out the fact that $X^3$ is a $U(1)$ bundle over $S^2$. This $U(1) $  bundle is nontrivial and may be interpreted as a magnetic monopole at the centre of $S^2$.  As we shall see, this monopole always has even integer charge.
As one of our prime objective is to adapt the above construction to the fuzzy version, we will mainly use group theoretic techniques, refraining from using any differential geometry.

This article is organised as follows.
In section 2 we review $Y^4$ and its base $X^3$, recalling their geometric properties and various symmetries. In section 3 we show that $X^3$ is a principal $U(1)$ bundle over $S^2$. The associated line bundles carry nontrivial monopole charge and describe topologically nontrivial configurations of a complex scalar field. In section 4 we show how these monopoles may be realized as arising from the Kaluza-Klein reduction of $X^3$ to $S^2$. This construction is very much in the spirit of \cite{Sorkin:1983ns, Gross:1983hb}. The fuzzy version is described in section 5. In section 5.1 we define the fuzzy conifold $Y^4_F$ and its base $X^3_F$ as certain subspaces of the Hilbert space of the 3--dimensional isotropic oscillator. In section 5.2 we show that the spin--1 matrices can be used to define a map $X^3_F\rightarrow S^2_F$. 
To describe the fuzzy monopoles we will adopt the stategy of  \cite{Grosse:1995jt}, directly constructing the sections of the fuzzy line bundles. Roughly speaking, these are  ``rectangular matrices'' that map fuzzy sphere of one size to another.

\section{ The Conifold $Y^4$ And Its Base $X^3 $}

The conifold $Y^4$  \cite{Parthasarathy:2001ss}  is a 4--dimensional manifold  embedded in $\mathbb{C}^3$ with 3--complex coordinates $z_\alpha$  $(\alpha=1,2,3)$ satisfying 
\begin{equation}
\mathcal{O}(z_\alpha)\equiv\sum_{\alpha=1}^3 z_\alpha^2 =0, \quad z_\alpha \in \mathbb{C}^3.
\label{conifold_1}
\end{equation}
It is an $O(3)$ symmetric smooth manifold with conical singularity  at a single point $z_\alpha=0 $, where the function $\mathcal{O}(z_\alpha)$ and its  derivatives vanish:
\begin{equation}
\mathcal{O}(z_\alpha)|_{z_\alpha=0} =0, \quad \left(\frac{\partial\mathcal{O}}{\partial z_\alpha}\right)_{z_\alpha=0} =0.
\end{equation}
The complex manifold $Y^4$ is the set of all lines passing through origin of $\mathbb{C}^3$ and hence a cone  
with a double singular point $z_\alpha=0$ as its  apex.

There is a scaling symmetry on $Y^4$: 
for any $\psi \in \mathbb{C}$ and any $z_\alpha$ obeying (\ref{conifold_1}), $\psi z_\alpha$ also solves  (\ref{conifold_1}). As we will see shortly,  under this transformation $z_\alpha \rightarrow \psi z_\alpha$,  the metric gets rescaled: $d\tilde{S}^2_{Y^4}  \rightarrow |\psi|^2 d\tilde{S}^2_{Y^4} $. The space has a $SO(3) \times U(1)$ symmetry with an isolated Calabi-Yau singularity  and the coordinates $z_\alpha$ transforms as vectors of $SO(3)$.

 The intersection of $Y^4$ with the unit sphere $S^5 \subset \mathbb{C}^3$ is called the base $X^3$.  It is  a smooth 3--dimensional manifold  devoid of any singularities and described by 
\begin{equation}
\mathcal{O}\equiv z_1^2+z_2^2+z_3^2=0, \quad   \bar{z}_1z_1+\bar{z}_2 z_2+\bar{z}_3z_3=1. 
\label{deformed_conifold}
\end{equation}
$X^3$ has $SO(3) \times U(1)$ symmetry and $Y^4$ is a cone over $X^3$.
The manifold $X^3$ can be parametrized as  \cite{Parthasarathy:2001ss}
\begin{eqnarray}
&&z_1=\frac{1}{\sqrt{2}}e^{2i\phi} \left(\cos^2\frac{\theta}{2} -e^{2i\xi} \sin^2\frac{\theta}{2}\right),\nonumber \\
&&z_2 = \frac{i}{\sqrt{2}}e^{2i\phi} \left(\cos^2\frac{\theta}{2} +e^{2i\xi} \sin^2\frac{\theta}{2}\right),\label{parametrization}\\
&&z_3= -\frac{1}{\sqrt{2}}e^{i(2\phi+\xi)} \sin\theta \nonumber
\end{eqnarray}
with  $0\leq \theta \leq \pi$, $-\pi\leq \xi \leq \pi$ and $0\leq \phi\leq 2\pi$.

To parametrize $Y^4$, we need to add a radial coordinate $r$ to the above parametrization. For $r \rightarrow \infty$ the metric on $Y^4$ can be written as in \cite{Pandozayas:2000sq}
\begin{equation}
ds^2_{Y^4} = dr^2 + r^2\left(a \left(d\phi + (1-\cos\theta) d\xi\right)^2 + b\left(d \theta^2 + \sin^2\theta d\xi^2\right)\right)
\end{equation}
where $a$ and $b$ are constants.
Demanding Ricci--flatness, we find that  $a=1/4=b$, giving us 
\begin{equation}
ds^2_{Y^4} = dr^2 + \frac{r^2}{4}\left( \left(d\phi + (1-\cos\theta) d\xi\right)^2 + \left(d \theta^2 + \sin^2\theta d\xi^2\right)\right).
\end{equation}
The scaling $z_\alpha \rightarrow \psi z_\alpha$ can be exploited to rescale the  $r\rightarrow\tilde{r} =\frac{r}{2}$. Then
\begin{equation}
ds^2_{Y^4} = 4d\tilde{r}^2 + \tilde{r}^2\left( \left(d\phi + (1-\cos\theta) d\xi\right)^2 + \left(d \theta^2 + \sin^2\theta d\xi^2\right)\right).
\end{equation}
The angular part of this metric can be identified as the metric on the base $X^3$:
\begin{eqnarray}
 ds^2_{X^3}\equiv ds^2=\left(d\theta^2 +\sin^2\theta d\xi^2\right)+\left( (1-\cos\theta) d\xi+ d\phi\right)^2.
\label{metric}
\end{eqnarray}

\section{ $X^3$  Is  a $U(1)$ Bundle Over  $S^2$ }
Let us define a map $\Pi:\mathbb{C}^3 \rightarrow \mathbb{R}^3$
\begin{equation}
y_i =z^\dagger I_i z, \quad \bar{y}_i = y_i,\quad \quad z=\left(\begin{array}{lll}
z_1\\z_2\\z_3
\end{array}\right)
\label{map}
\end{equation}
where $i=1,2,3$ and $I_i$ are $3 \times 3$ matrices 
\begin{equation}
(I_i)_{\alpha\beta} = -i\epsilon _{i\alpha\beta}, \quad\quad \mathrm{where}\quad \alpha, \beta=1,2,3
\end{equation}
 These are the generators of the $SO(3)$ algebra in the fundamental representation:
\begin{eqnarray}I_1=\left(
\begin{array}{llll}
0&0&0\\
0&0&-i\\
0&i&0
\end{array}\right),
 \quad 
I_2=\left(
\begin{array}{llll}
0&0&i\\
0&0&0\\
-i&0&0
\end{array}\right) 
\quad\mathrm{and} \quad
I_3=\left(
\begin{array}{llll}
0&-i&0\\
i&0&0\\
0&0&0
\end{array}\right)
\label{so3_matrices}
\end{eqnarray}
satisfying 
\begin{equation}
[I_i,I_j]=i\epsilon _{ijk}I_k ,
\end{equation}
with the Casimir
\begin{equation}
\sum_{i=1}^3I_iI_i =\left( \begin{array}{ccc}
2&0&0 \\0&2&0\\
0&0&2
\end{array}\right).
\label{Casimir1}
\end{equation}
 These $y_i $ satisfy
\begin{equation}
\sum_{i=1}^3 y_i y_i  =\left(\sum_{\alpha=1}^3 \bar{z}_\alpha z_\alpha\right)^2 -\sum_{\alpha,\beta=1}^3\bar{z}_\alpha^2z_\beta^2=\left(\sum_{\alpha=1}^3 \bar{z}_\alpha z_\alpha\right)^2 -\bar{\mathcal{O}}\mathcal{O}.
\label{casimir22}
\end{equation}
So if $z_\alpha\in X^3$, then
\begin{equation}
\sum_{i=1}^3 y_iy_i  =1.
\end{equation}
This space is the unit sphere $S^2$   and thus (\ref{map}) is a map  $ \Pi: X^3  \rightarrow  S^2$.
Using the (\ref{parametrization}), we can explicitly write $\Pi$ as
\begin{equation}\left.\begin{array}{lll}
y_1=&-i\left(\bar{z}_2z_3-\bar{z}_3z_2\right)=&\sin\theta \cos\xi,\\
y_2=&-i\left(\bar{z}_3z_1-\bar{z}_1z_3\right) =& \sin\theta\sin\xi ,\\
y_3=&-i\left(\bar{z}_1z_2-\bar{z}_2z_1\right)=&\cos \theta,\end{array}\right.
\label{map_coordinates}
\end{equation}
where $0\leq \theta \leq \pi$ and $-\pi \leq \xi \leq \pi$.

$X^3$ has a $U(1)$ symmetry  as (\ref{deformed_conifold}) is invariant under  $z_\alpha \rightarrow e^{i\lambda} z_\alpha$. For $z_\alpha \in X^3$,  $\Pi$ is also invariant under this transformation. It is evident from (\ref{map_coordinates}) that the sphere $S^2$ is independent of  $\phi$.  This means $\Pi$ maps circles $S^1$ on $X^3$ to points on $S^2$.  $X^3$ is thus a  $U(1)$ bundle over  $S^2$:
\begin{eqnarray}
U(1) &\rightarrow & X^3 \nonumber \\ 
&& \downarrow \label{fibration}\\ 
&& S^2. \nonumber
\end{eqnarray}
The angles $\theta$ and $\xi$ are the coordinates of $S^2$ and the fibre  cordinate is $\phi$.

It is useful to explicitly describe the coordinate charts that we will use on $S^2$.
To this end, we define the complex functions 
\begin{eqnarray}
&&w^N_1=\frac{1}{\sqrt{2}} \left(\cos^2\frac{\theta}{2} -e^{2i\xi} \sin^2\frac{\theta}{2}\right), \nonumber\\
&& w^N_2 = \frac{i}{\sqrt{2}}\left(\cos^2\frac{\theta}{2} +e^{2i\xi} \sin^2\frac{\theta}{2}\right), \label{coordinate_chart1}\\
&& w^N_3= -\frac{1}{\sqrt{2}}e^{i\xi} \sin\theta. \nonumber 
\end{eqnarray}
These functions are well--defined at all points on $S^2$ except for $\theta= \pi$ (the South Pole $S$). It is easy to see that  (\ref{coordinate_chart1}) is obtained by setting $\phi=0$ in (\ref{parametrization}).  Let us  denote this coordinate chart as $U_N$.
To describe the coordinate chart $U_S$ that includes the South Pole
we set $\phi =-\xi$ in (\ref{parametrization}) to obtain 

\begin{eqnarray}
&&w^S_1=\frac{1}{\sqrt{2}} \left(e^{-2i\xi} \cos^2\frac{\theta}{2} - \sin^2\frac{\theta}{2}\right), \nonumber\\
&& w^S_2 = \frac{i}{\sqrt{2}}\left(e^{-2i\xi} \cos^2\frac{\theta}{2} +\sin^2\frac{\theta}{2}\right), \label{coordinate_chart2} \\
&&w^S_3= -\frac{1}{\sqrt{2}}e^{-i\xi} \sin\theta. \nonumber 
\end{eqnarray}
These functions are well--defined at all points on $S^2$ except at $\theta=0$ (the North Pole $N$).  On the overlapping region $U_N \cap U_S$, 
\begin{equation}
w^N_\alpha = e^{2i\xi} w^S_\alpha.
\end{equation}
 It is important to note that the (\ref{coordinate_chart1}) and (\ref{coordinate_chart2}) are not the standard stereographic projection maps.

Now let us describe the topologically non--trivial configurations of a complex scalar field on this $S^2$.
A complex scalar field on $S^2_N\equiv S^2-\{S\}$ is a function of $w^N_\alpha$
\begin{equation}
\Phi_N = \sum c_{n_1 n_2 n_3n'_1n'_2n'_3} (\bar{w}^N_1)^{n'_1} (\bar{w}^N_2)^{n'_2}(\bar{w}^N_3)^{n'_3} (w^N_1)^{n_1} (w^N_2)^{n_2}(w^N_3)^{n_3}
\end{equation}
while on the other patch $S^2_S\equiv S^2-\{N\}$, it is a function of $w^S_\alpha$
\begin{equation}
\Phi_S = \sum c_{n_1 n_2 n_3 n'_1n'_2n'_3} (\bar{w}^S_1)^{n'_1} (\bar{w}^S_2)^{n'_2} (\bar{w}^S_3)^{n'_3} (w^S_1)^{n_1} (w^S_2)^{n_2}(w^S_3)^{n_3}.
\end{equation}
If $k= n'_1+n'_2+n'_3-n_1-n_2-n_3=$ fixed, then
 in the region $S^2_N\cap S^2_S$, $\Phi_N $ is related to $\Phi_S$ as
\begin{equation}
\Phi_N=e^{2i (n'_1+n'_2+n'_3-n_1-n_2-n_3) \xi}\Phi_S = e^{i \kappa \xi}\Phi_S,\quad \kappa= 2k.
\end{equation}
We recognise the phase in the above equation as the  gauge transformation relating $\Phi_N$ and $\Phi_S$. 
 This gauge transformation arises from a gauge field $A_\mu$ with
\begin{eqnarray}
&A^N_\mu = - i \frac{\kappa}{2} \bar{w}^N_\alpha \left(\partial_\mu w^N_\alpha\right) \quad \mathrm{on}\,\,\, S^2_N,\nonumber\\
&A^S_\mu =- i \frac{\kappa}{2} \bar{w}^S_\alpha \left(\partial_\mu w^S_\alpha\right)\quad \mathrm{on}\,\,\, S^2_S,\\
&\mathrm{and}\,\,\, \,\,\,\, A^N_\mu = A^S_\mu + i e^{i\kappa \xi}\left(\partial_\mu e^{-i\kappa \xi}\right)\quad \mathrm{on}\,\,\, S^2_N\cap S^2_S \nonumber
\end{eqnarray}
where $\mu = \theta, \xi$.
Explicit computation gives
\begin{equation}
A^N_\theta=0,\quad A^N_\xi = \frac{\kappa}{2}(1-\cos \theta); \quad A^S_\theta=0,\quad A^S_\xi = -\frac{\kappa}{2}(1+\cos \theta).
\end{equation}
The connection one-forms are
\begin{equation}
A^N= A^N_\theta d\theta + A^N_\xi d\xi = \frac{\kappa}{2}(1-\cos \theta) d\xi; \quad A^S= A^S_\theta d\theta + A^S_\xi d\xi = -\frac{\kappa}{2}(1+\cos \theta) d\xi
\end{equation}
In the overlapping region $S^2_N \cap S^2_S$, $A_N$ and $A_S$ are related as $A^N-A^S= \kappa d\xi$ where $\kappa$ is even integer.

We denote by $\mathcal{H}_\kappa$ the space of these complex scalar fields with a fixed $\kappa$. The elements $\Phi$ of $\mathcal{H}_\kappa$ are eigenfunctions of the operator $K_0$ with eigenvalue $\frac{\kappa}{2}$:
\begin{equation}
K_0 \Phi\equiv\sum_{\alpha=1}^3 \left(\bar{w}_{\alpha}\frac{\partial}{\partial \bar{w}_{\alpha}} - w_{\alpha}\frac{\partial}{\partial w_{\alpha}}\right) \Phi = \frac{\kappa}{2}\Phi.
\end{equation}
We will therefore call this operator as the topolological charge operator.
 $\mathcal{H}_\kappa$ is the linear  space of sections $\Phi$  with has a topological charge $\kappa$.

The differential operators $J_i=-i\epsilon_{ijk}y_j\frac{\partial \,\,\,\,}{\partial y_k}$ can be written in terms of $w_\alpha$:
\begin{eqnarray}
J_1=-i \left(\bar{w}_2\frac{\partial}{\partial \bar{w}_3}-w_3\frac{\partial}{\partial w_2}-\bar{w}_3\frac{\partial}{\partial \bar{w}_2}+w_2\frac{\partial}{\partial w_3}\right) \nonumber \\
J_2=-i \left(\bar{w}_3\frac{\partial}{\partial \bar{w}_1}-w_1\frac{\partial}{\partial w_3}-\bar{w}_1\frac{\partial}{\partial \bar{w}_3}+w_3\frac{\partial}{\partial w_1}\right) \\
J_3=-i \left(\bar{w}_1\frac{\partial}{\partial \bar{w}_2}-w_2\frac{\partial}{\partial w_1}-\bar{w}_2\frac{\partial}{\partial \bar{w}_1}+w_1\frac{\partial}{\partial w_2}\right).  \nonumber
\end{eqnarray}
The $J_i$'s act on $\mathcal{H}_\kappa$ and map $\mathcal{H}_\kappa \rightarrow \mathcal{H}_\kappa$.
In $\mathcal{H}_\kappa$, $J_i$'s satisfy the $SU(2)$ algebra
\begin{equation}
[J_i,J_j]=i\epsilon_{ijk}J_k .
\label{adjoint_su2}
\end{equation}
The topological charge operator $K_0$ commutes with $J_i$:
\begin{equation}
[K_0,J_i]=0.  
\end{equation}
In $\mathcal{H}_\kappa$  we can choose the eigenfunctions of $J_3$ and $J_i j_i$ as a basis to expand  any $\Phi \in \mathcal{H}_\kappa$.

Now let us construct these basis functions. 
The function $h=(\bar{w}_1+i \bar{w}_2)^{l} (w_1+i w_2)^{n}$ is an element of $\mathcal{H}_\kappa$ with $\kappa =2(l-n)$ as 
\begin{equation}
K_0 h = (l-n) h.
\end{equation}
This function $h$ satisfies
\begin{equation}
J_+ h =0, \quad  J_3 h= (l+n) h  
\label{highest_weight}
\end{equation}
and so $h$ is the highest weight vector of the  (\ref{adjoint_su2}) with $j=(l+n)$.  We denote this highest weight vector $h$ by $\Phi^j_{\kappa, m=j}$, from which the lower weight vectors can be obtained by repeated application of the $J_-$: 
\begin{equation}
(J_-)^{(j-m)} \Phi^j_{\kappa,m=j} = N^{j}_{\kappa,m}\Phi^j_{\kappa,m}, \quad -j\leq m\leq j.
\end{equation}
The constants $ N^{j}_{\kappa,m}$ can be evaluated explicitly
 but are unnecessary for our purposes, and we will not do so here.

The value of $j$ is given by (\ref{highest_weight}) 
\begin{equation}
j\equiv l+n = \frac{\kappa }{2}+2n.
\end{equation}
As $l$ and $n$  take  values $0,1,2,3\dots$ and as $\kappa$ can have only even integer value, while $j$  takes all integer values greater that $\frac{\kappa}{2}$ : 
\begin{equation}
j=\frac{\kappa}{2},\frac{\kappa}{2}+2,\frac{\kappa}{2}+4,\ldots
\end{equation}

The set  $\{\Phi^j_{\kappa,m}\}$ spans $\mathcal{H}_\kappa$ 
and any  element $\Phi$ of $\mathcal{H}_\kappa$ can be expressed as 
\begin{equation}
\Phi = \sum_{j=\frac{\kappa}{2}}^{\infty} \sum_{m=-j}^{j} c^j_{\kappa, m}\Phi^j_{\kappa,m},\quad \quad c^j_{\kappa,m} \in \mathbb{C}.
\end{equation}
These elements of $\mathcal{H}_\kappa$ are identified as sections of the line bundle with topological charge $\kappa$ (= even integer).

\section{Monopoles From Kaluza-Klein Reduction $X^3 \rightarrow S^2$}

Interestingly, the principal fibre bundle of the previous discussion can be obtained by the Kaluza-Klein reduction of  the metric on  $X^3$. 
The metric (\ref{metric}) can be written as $ds^2= \tilde{g}_{ab} d\eta^a d\eta^b$ where
\begin{equation}\tilde{g}_{ab}=\left(
\begin{array}{cccc}
1&0&0 \\
0&\sin^2 \theta+4 \sin^4\frac{\theta}{2}&2\sin^2\frac{\theta}{2}\\
0&2\sin^2\frac{\theta}{2}&1
\end{array}\right);\quad \eta^1=\theta, \eta^2=\xi, \eta^3=\phi.
\label{metric12}
\end{equation}
It is an Einstein metric since
\begin{eqnarray}
\tilde{R}_{ab}=\frac{1}{2}\tilde{g}_{ab}
\end{eqnarray}
and it solves Einstein equations with a positive cosmological constant.

The metric (\ref{metric12}) already has the convenient   Kaluza-Klein form
\begin{equation}\tilde{g}_{ab}=\left(
\begin{array}{cccc}
g_{\mu\nu}+g A_\mu A_\nu &g A_\mu\\
g A_\nu & g
\end{array}\right), \quad 
g_{\mu\nu}=\left(
\begin{array}{cccc}
1&0\\
0& \sin^2 \theta
\end{array}\right),
\label{kk_metric}
\end{equation}
with one extra compact dimension $\phi$ and the dilaton $g$ set to 1.
Inspecting (\ref{metric12}) immediately tells us that  $g_{\mu\nu}$ is  the metric on $S^2_N$ and the gauge fields on this local patch are $A^N_\theta=0$, $A^N_\xi=2 \sin^2 \frac{\theta}{2}=(1-\cos \theta)$. The field strength $F^N= dA^N$ is 
\begin{equation}
F^N_{\theta \xi}=  \frac{\partial A^N_\xi}{\partial \theta}-\frac{\partial A^N_\theta}{\partial \xi} = \sin \theta.
\label{mag_rad}
\end{equation}
This corresponds to a  constant radial magnetic field $B^N_r= * F^N$  and hence   a monopole of charge $\kappa= 2$ at the center of the sphere.

Similar computations on $S^2_S$ give us 
$A^S_\theta=0$, $A^S_\xi= -(1+\cos \theta)$.  $A_\xi^N$ and $A_\xi^S$ are related by the gauge transformation $A^N_\xi=A^S_\xi +i g\partial_\xi g^{-1}$, where $g=e^{2i\xi} \in U(1)$.
This is a simplified version of the Kaluza-Klein monopole (for the standard KK monopole see \cite{Gross:1983hb, Sorkin:1983ns}) . 

Notice that the metric on $X^3$ is in the KK-form. Further  both $X^3$ and $S^2$ are Einstein manifolds. This observation provides us with a handle for generalizing this construction to other manifolds and dimensions. Consider the Kaluza-Klein reduction of $\mathcal{M}^{d+1}$ to $\mathcal{M}^{d}$ where both $\mathcal{M}^{d+1}$ and $\mathcal{M}^{d}$ are compact Einstein manifolds.
As we shall see below, the ``Einstein condition'' (the Ricci tensor is proportional to the metric) leads to stringent conditions on the gauge field. When $\mathcal{M}^{d}$ is $S^2$ it also leads to a relation between the monopole charge, size of $S^2$, the dilaton $g$ and the cosmological constant.

Consider  a $(d+1)$--dimensional manifold $\mathcal{M}^{d+1}$ with metric $\tilde{g}_{ab}$ in the form
(\ref{kk_metric}). The $g_{\mu\nu}$ is identified as the metric on a $d$--dimensional manifold $\mathcal{M}^{d}$. There is one extra-dimension which we assume is compact, and that $\tilde{g}_{ab}$ is  independent of the coordinate of the extra dimension (Kaluza's cylinder condition). Let $\tilde{g}_{d+1,d+1} \equiv g$ be a constant. The components of the Ricci tensor are 
\begin{eqnarray}
&&\tilde{R}_{d+1,d+1} = \frac{g^2}{4} F^{\sigma\beta}F_{\sigma\beta},
 \quad \tilde{R}_{\mu, d+1}=\frac{g^2}{4} F^{\sigma\beta}F_{\sigma\beta} A_\mu=\tilde{R}_{d+1,\mu},\nonumber \\
&&\tilde{R}_{\mu\nu}= R_{\mu\nu} +\frac{g^2}{4} F^{\sigma\beta}F_{\sigma\beta} A_\mu A_\nu -\frac{g}{4}F^{\sigma\beta}\left(g_{\mu \beta}F_{\sigma\nu}+g_{\nu \beta}F_{\sigma\mu}\right)
\end{eqnarray}
where $F_{\sigma\beta}=\partial_\sigma A_\beta -\partial_\beta A_\sigma$ and $F^{\sigma\beta}=g^{\rho\sigma}g^{\alpha\beta}F_{\rho\alpha}$.

Since both  $\mathcal{M}^{d+1}$ and $\mathcal{M}^{d}$ are Einstein manifolds (with different cosmological constants!)
\begin{equation}
\tilde{R}_{ab}= c\tilde{g}_{ab}, \quad R_{\mu\nu}= c_0 g_{\mu\nu}.
\end{equation}
It is easy to see that the gauge fields must satisfy the conditions
\begin{eqnarray}
&&F^{\sigma\beta}F_{\sigma\beta} = \frac{4c}{g}, \label{cond3}\\
&& F^{\sigma\beta}\left(g_{\mu \beta}F_{\sigma\nu}+g_{\nu \beta}F_{\sigma\mu}\right) = \frac{4}{g} (c_0 - c)g_{\mu\nu}.
\label{cond2}
\end{eqnarray}
Multiplying (\ref{cond2}) by $g^{\mu\nu}$ and summing over $\mu$ and $\nu$,  we get 
\begin{equation}
F^{\sigma\beta}F_{\sigma\beta} =\frac{2}{g}(c_0 - c)d
\end{equation}
 yielding 
\begin{equation}
c_0= \frac{d+2}{d}c.
\label{reln1}
\end{equation}
Moreover, as $\mathcal{M}^{d}$ is compact, (\ref{cond3}) implies that  the electromagnetic action 
\begin{equation}
S_{EM} \propto \int d(Vol) F^{\sigma\beta}F_{\sigma\beta}=\frac{4c}{g} Volume
\end{equation}
is finite.

Let us specialize to the case when $\mathcal{M}^{d+1} $ is 3--dimensional and $\mathcal{M}^{d}=S^2$. For example $\mathcal{M}^{d+1}$ could be $S^3$ (the standard Hopf fibration) or $X^3$ (our (\ref{map_coordinates})).
Then (\ref{reln1}) gives $c_0=2c$ and the metric on $S^2$ is 
\begin{equation}
g_{\mu\nu}=\left(\begin{array}{lll}
R^2&0\\
0& R^2 \sin^2 \theta
\end{array}\right), \quad R=\left\{ \begin{array}{lll}
\frac{1}{2} \,\,\, \mathrm{for}\,\,\, S^3 \rightarrow S^2\\ \\
1 \,\,\, \mathrm{for}\,\,\, X^3 \rightarrow S^2.
\end{array}\right.
\end{equation}
 
>From ({\ref{cond2}), we can write
\begin{equation}
F_{12}= \sqrt{\frac{2c}{g} \frac{g_{11}}{g^{22}}}=R^2 \sqrt{\frac{2c}{g}}\sin \theta.
\end{equation}
So, there is  a radial magnetic field which corresponds to a magnetic monopole of charge $\kappa = 2R^2\sqrt{\frac{2c}{g}}$ at the centre of the sphere. For $S^3$ this gives $F_{12}=\frac{1}{2} \sin \theta$ and $\kappa=1$ where as for $X^3$ we get $F_{12}= \sin \theta$ and $\kappa=2$.

\section{Fuzzy Fibre Bundle}

A large number of fuzzy spaces arise as quantized symplectic manifolds that are co-adjoint orbits of Lie groups \cite{perelomov,kirillov}. We will not attempt to provide a mathematically precise definition of a fuzzy space here (a universally acceptable definition does not seem to exist in the known literature), rather our aim is to present a description of the fuzzy conifold and its monopole bundle in the spirit of \cite{Grosse:1995jt}.

 Often, such spaces also have a description in terms of  bosonic oscillators $\hat{a}_\alpha,\hat{a}_\alpha^\dagger$ (or more precisely, specific operator functions of $\hat{a}_\alpha,\hat{a}_\alpha^\dagger$) (see \cite{Balachandran:2005ew}).
Such spaces are topics of interest to physicists and mathematicians especially in context of finite dimensional approximations of quantum field theories. Fuzzy spaces have encrypted topological information and  provide  new insights into the discrete analogues of  instantons, solitons, monopoles and the like (for example see \cite{ {Hoppe:1982},{Madore:1991bw},{Grosse:1992bm}, {Baez:1998he},{Balachandran:1999hx},{Vaidya:2001rf},{Nekrasov:1998ss},{Acharyya:2011bx}}). 

 The fuzzy  $n$--dimensional complex plane $\mathbb{C}^n_F$ is represented by $n$ independent  bosonic oscillators
\begin{equation}
[\hat{a}_\alpha, \hat{a}_\beta^\dagger]=\delta_{\alpha\beta},\quad  \,\,\, \, \alpha,\beta=1,2,3\ldots n.
\end{equation}
These operators act on the Hilbert space of the $n$ bosonic oscillators. Restricting   to appropiate subspaces of this Hilbert space gives us other fuzzy spaces like  $S^{2n-1}_F$, $\mathbb{CP}_F^n$ and so on \cite{Balachandran:2005ew,Balachandran:2001dd, Dolan:2006tx}.

 For the $n=2$ the Fock space of states on which these operators operate is spanned by the states $|n_1,n_2\rangle$ -- the eigenstates of the number operator $N=\hat{a}_1^\dagger \hat{a}_1  + \hat{a}_2^\dagger \hat{a}_2$. In the subspaces $ f_n = \{  |n_1,n_2\rangle, n_1+n_2 =n=\mathrm{fixed} \}$, the number operator $N$ takes a constant value and the restriction to this subspace defines the fuzzy manifold $S^3_F$.
The Jordan-Schwinger realization of the $SU(2)$ algebra  is an operator map $L_i = \frac{1}{2} \hat{a}_\alpha^\dagger (\sigma_i)^{\alpha\beta}\hat{a}_\beta$ such that
\begin{equation}
[L_i,L_j]=i \epsilon_{ijk}L_k, \quad \sum_{i=1}^3 L_iL_i = \frac{N}{2}\left(\frac{N}{2}+1\right).
\label{fuzzy_s2}
\end{equation}
In the subspace $f_n$, the Casimir $L_i L_i$  is  fixed and resultant fuzzy space is $S^2_F$.
This map $S_F^3 \rightarrow S_F^2$  is the noncommutative Hopf fibration.

Though standard differential geometric tools are unavailable for  these discrete spaces, much topological  information  can still be extracted indirectly by studying the  the complex line bundles.  A particularly simple approach has been developed by \cite{Grosse:1995jt}, where the group action of $SU(2)$ is used to identify the fuzzy line bundles. It has gained some special attention due to its simplicity and the lucid connection of the approach with the continuous case.

The idea of noncommutative Hopf map can be generalized to higher  dimensional noncommutative spaces. In particular, we are interested in  a``Hopf-like" construction which relates the fuzzy conifold $Y^4_F$ with $S_F^2$. We show below that the techniques of \cite{Grosse:1995jt} can be adapted to describe the fuzzy fibration $X^3_F \rightarrow S_F^2$, construct the corresponding line bundles and identify the monopole charges.

\subsection{The Fuzzy Conifold $Y_F^4$ And The Base $X_F^3$ }
$\mathbb{C}^3_F$  is described by the algebra of  three independent oscillators
\begin{equation}
[\hat{a}_\alpha, \hat{a}_\beta^\dagger]=\delta_{\alpha\beta}, \quad\quad \mathrm{where} \quad \alpha,\beta=1,2,3.
\end{equation}
which acts on a space $\mathcal{F}$  spanned by the eigenstates of the {\it number} operators $\hat{N}_\alpha \equiv \hat{a}_\alpha^\dagger \hat{a}_\alpha$:
\begin{equation}
\mathcal{F}= Span\left\{|n_1,n_2, n_3 \rangle: n_\alpha=0,1,2.....\right\}.
\end{equation}
The total number operator is $$\hat{N} \equiv\sum_{\alpha=1}^3\hat{N}_\alpha.$$

In analogy with  (\ref{conifold_1})
let us define the operator $\hat{\mathcal{O}}$ as
\begin{equation}
\hat{\mathcal{O}} \equiv \hat{a}_1^2 +\hat{a}_2^2+\hat{a}_3^2
\end{equation}
which has as its kernel 
\begin{equation}
\ker(\hat{\mathcal{O}}) = \left\{|n_1,n_2, n_3 \rangle: \,\, \hat{\mathcal{O}}|n_1,n_2, n_3 \rangle=0 \right\}.
\end{equation}

The usual conifold (\ref{conifold_1}) is defined by the commutative algebra of polynomial functions of $z_\alpha$, subject to the condition 
$\sum_\alpha z_\alpha^2=0$. It is tempting to define its noncommutative analogue as the algebra of the polynomial functions of $\hat{a}_\alpha$, subject to the condition $\sum_\alpha \hat{a}_\alpha^2=0$. This algebra as it stands is a commutative algebra. 
However $\hat{a}_\alpha$'s are not ordinary complex variables, rather they are infinite dimensional operators.  Under the $\ast$--operation (where $(\hat{a}_\alpha)^\ast \equiv \hat{a}_\alpha^\dagger$) this algebra as a star algebra is noncommutative. 
We will therefore interpret the condition  $\sum_\alpha \hat{a}_\alpha^2=0$ as a constraint on the set of admissible states in the bosonic Hilbert space of $\hat{a}_\alpha$'s.

Thus our operational  definition of the fuzzy  conifold $Y_F^4$ is the restriction of the action of the operators $\hat{a}_{\alpha}$  (and polynomial operator functions of $\hat{a}_\alpha$) to $\ker(\hat{\mathcal{O}})$. Defining a fuzzy space by such a restriction is not new. For example, in the paragraph preceeding (\ref{fuzzy_s2}),  $S_F^2$ is defined as the restriction of $L_i$ (an operator function of $a_\alpha$) to the subspace $f_n$ which are the finite dimensional representation of  $SU(2)$.

Also, it is easy to see why our operational definition of $Y_F^4$ is appropriate. Let $|z_1,z_2,z_3 \rangle$ be the  standard coherent states of 3--dimensional oscillator
\begin{equation}
|z_1,z_2,z_3 \rangle=N(z_\alpha, \bar{z}_\alpha) e^{z_\alpha \hat{a}_\alpha^\dagger}|0,0,0\rangle.
\label{coherent_states}
\end{equation}
Then 
\begin{equation}
\hat{\mathcal{O}} |z_1,z_2,z_3 \rangle = \left(z_1^2+z_2^2+z_3^2\right)|z_1,z_2,z_3 \rangle
\end{equation}
This implies that if $|z_1,z_2,z_3 \rangle \in \ker (\hat{\mathcal{O}})$, the complex numbers $z_\alpha$ obey the conifold condition (\ref{conifold_1}).

The operators 
\begin{equation}
\hat{\chi}_\alpha = \hat{a}_\alpha \frac{1}{\sqrt{\hat{N}}}, \quad \hat{\chi}_\alpha^\dagger =  \frac{1}{\sqrt{\hat{N}}}\hat{a}_\alpha^\dagger
\end{equation}
satisfy 
\begin{equation}
\hat{\chi}_\alpha^\dagger \hat{\chi}_\alpha = 1.
\label{five_sphere}
\end{equation}
We exclude the state $|0,0,0 \rangle$ from the domain of $\hat{\chi}_\alpha$ so that $\hat{\chi}_\alpha$ is well-defined.
Then (\ref{five_sphere}) gives us the fuzzy 5--sphere $S_F^5$. 
The operator
\begin{equation}
\hat{\mathcal{O}}' \equiv\sum_{\alpha=1}^3\hat{ \chi}_\alpha^2 = \frac{1}{\sqrt{(\hat{N}+1)(\hat{N}+2)}} \hat{\mathcal{O}}
\end{equation}
obviously vanishes on $\ker(\hat{\mathcal{O}})$. The restriction of  $\hat{\chi}_\alpha$ to $\ker(\hat{\mathcal{O}})$  defines for us $X^3_F$, the fuzzy version of $X^3$. One may think of $X^3_F$ as the intersection of $Y^4_F$ with $S_F^5$.

The continuum $X^3$ can be recovered in the limit $\hat{N} \rightarrow \infty$ of $X_F^3$. It is simplest to see this using the coherent states (\ref{coherent_states}) but we will skip the details here.

\subsection{Fuzzy Two--Sphere $S_F^2$ And The Noncommutative Fibre Bundle}

With the  matrices (\ref{so3_matrices}), we can write the analogue of  the map (\ref{map}):
\begin{eqnarray}
\hat{y}_i = \hat{\chi}^\dagger I_i \hat{\chi}=\frac{1}{\hat{N}} \hat{L}_i, \quad  \mathrm{where}\quad\hat{L}_i = \hat{a}^\dagger_\alpha (I_i)_{\alpha\beta} \hat{a}_\beta \quad \mathrm{and}\quad \hat{\chi}=\left(\begin{array}{lll}
\hat{\chi}_1 \\
\hat{\chi}_2\\
\hat{\chi}_3
\end{array}\right).
\label{map2}
\end{eqnarray}
The $\hat{L}_i$'s satisfy 
\begin{eqnarray}
[\hat{L}_i, \hat{L}_j]=i\epsilon_{ijk} \hat{L}_k, \quad [\hat{L}_i, \hat{N}]=0.
\label{su2_1}
\end{eqnarray}
The Casimir can be conveniently expressed as
\begin{equation}
\hat{C}= \hat{L}_i\hat{L}_i =\hat{N}\left(\hat{N}+1\right) -\hat{\mathcal{O}}^\dagger \hat{\mathcal{O}}.
\label{casimir21}
\end{equation}
As $\hat{L}_i$ and $\hat{N}$ commute, it is easy to see that 
\begin{equation}
\hat{y_i}\hat{y_i}= \frac{1}{\hat{N}^2} \hat{L}_i \hat{L}_i=\left(1+\frac{1}{\hat{N}}\right) - \frac{1}{\hat{N}^2}\hat{\mathcal{O}}^\dagger \hat{\mathcal{O}}.
\end{equation}
In $\ker (\hat{\mathcal{O}})$, $\hat{C}$ and $\hat{y_i}\hat{y_i}$ are simply  
\begin{equation}
\hat{C}|_{\ker(\hat{O})} =\hat{N}\left(\hat{N}+1\right) , \quad \quad \hat{y_i}\hat{y_i}|_{\ker(\hat{O})}=\left(1+\frac{1}{\hat{N}}\right).
\end{equation}

It is convenient to decompose $\mathcal{F}$ into subspaces $\mathcal{F}_n$ in which $\hat{N} $ takes a fixed value $n$:
\begin{equation}
\mathcal{F}_n= \left\{|n_1,n_2, n_3 \rangle: n_1+n_2+n_3=n \right\}, \quad \mathcal{F}=\oplus_n \mathcal{F}_n.
\end{equation}
The dimension $d_n$ of $\mathcal{F}_n$  is  $\frac{(n+1)(n+2)}{2}$.

$\tilde{\mathcal{F}}_n$ is the subspace of $\mathcal{F}_n$ defined as
\begin{equation}
\tilde{\mathcal{F}}_n= \mathcal{F}_n \cap \ker(\hat{\mathcal{O}}).
\end{equation}
It has the nice property that both $\hat{\mathcal{O}}$ vanishes and the value of $\hat{N}$ is fixed. So in $\tilde{\mathcal{F}}_n$
\begin{equation}
\hat{y}_i\hat{y}_i = \left(1+\frac{1}{n}\right)\mathbb{1}.
\end{equation}
When restricted to $\tilde{\mathcal{F}}_n$, the Casimir $\hat{C}$ takes the fixed value $n\left(n+1\right)$ and $\tilde{\mathcal{F}}_n$ is the carrier space for the $(2n+1)$ dimensional UIR of the $SU(2)$. As $n$ takes  integer values, only the odd dimensional representations occur in this construction.  This construction was first done in \cite{diptiman,diptiman2} in the context of ferromagnetism.

Thus the algebra generated by $\hat{y}_i$'s restricted to $\tilde{\mathcal{F}}_n$ is the fuzzy two--sphere $S_F^2$, and  (\ref{map2}) is a map $X_F^3 \rightarrow S^2_F$. 
The $n\rightarrow\infty$ is the commutative limit and in this limit as $\hat{y}_i \hat{y}_i \rightarrow 1$, we recover  $S^2$.

We will  now use the  $SU(2)$ group theory  to construct the noncommutative fibre bundles on this $S_F^2$. Our strategy will be  similar to the one in \cite{Grosse:1995jt}.

Let $\mathcal{H}_{nl} $ be the space of linear operators $\Phi$ which map $\tilde{\mathcal{F}}_n$ to $\tilde{\mathcal{F}}_l$: 
\begin{equation}
\Phi : \tilde{\mathcal{F}}_n \rightarrow \tilde{\mathcal{F}}_l,\quad \Phi \in \mathcal{H}_{nl}.
\end{equation}
The operators $\Phi$ can be represented by rectangular matrices of size $(2l+1) \times (2n+1)$. 
 The spaces $ \mathcal{H}_{nn}$ are $(2n+1)^2$ dimensional noncommutative algebras $\mathcal{A}_n$ which map $\tilde{\mathcal{F}}_n \rightarrow \tilde{\mathcal{F}}_n$.
The space $ \mathcal{H}_{nl}$ is a noncommutative bimodule: it is left $\mathcal{A}_l$--module and a right $\mathcal{A}_n$--module.

Rotations are generated in  $\mathcal{H}_{nn}$ by the adjoint action of $\hat{L}^{(n)}_i$:
\begin{equation}
Ad(\hat{L}_i) \Phi \equiv \hat{\mathcal{L}}_i \Phi\equiv [\hat{L}^{(n)}_i, \Phi], \quad \Phi \in  \mathcal{H}_{nn}
\end{equation}
and 
\begin{equation}
[\hat{\mathcal{L}}_i, \hat{\mathcal{L}}_j] =i \epsilon_{ijk} \hat{\mathcal{L}}_k.
\label{angular_mom}
\end{equation}
Since  $\tilde{\mathcal{F}}_n$ is the carrier space for the $(2n+1)$--dimensional UIR, the $\hat{L}^{(n)}_i$ above are the usual  $(2n+1)\times (2n+1) $ matrices.

On the bimodules $ \mathcal{H}_{nl}$, the generators of the $SU(2)$  algebra acts as 
\begin{equation}
\hat{\mathcal{L}}_i\Phi = \hat{L}_i^{(l)}\Phi - \Phi \hat{L}_i^{(n)}.
\label{su2_bimodule}
\end{equation}
 This action of  $SU(2)$ corresponds to the direct product  $l\otimes n$ of the  two UIRs $l$ and $n$. The elements of  $\mathcal{H}_{nl}$ can therefore be expanded in terms of the eigenfunctions of $\hat{\mathcal{L}}_3$ and $\hat{\mathcal{L}}_i\hat{\mathcal{L}}_i$ belonging to the irreducible representations in the decomposition of $l\otimes n$:
\begin{equation}
l\otimes n = |l-n|\oplus|l-n|+1\oplus\ldots\oplus (l+n). 
\end{equation}
We denote the minimum and the maximum values in this series as
\begin{equation}
|l-n|\equiv\frac{\kappa}{2}, \quad l+n\equiv J.
\end{equation}

Below we construct these basis functions explictly. The sections of the fuzzy line bundle can be expanded in terms of these basis functions. 

The operator 
\begin{equation}
\hat{h} =N_{\tilde{l}\tilde{n}} (\hat{\chi}_1^\dagger+i\hat{\chi}_2^\dagger)^{\tilde{l}}(\hat{\chi}_1+i\hat{\chi}_2)^{\tilde{n}}, \quad N_{\tilde{l}\tilde{n}} = \mathrm{constant}
\end{equation}
 is an element of $\mathcal{H}_{nl}$ if $0\leq \tilde{n} \leq n $ and $\tilde{l}-\tilde{n} = l-n\equiv \frac{\kappa}{2}.$  
Let us define a new set of oscillators $\{\hat{A}_1, \hat{A}_2, \hat{A}_3\}$ as
\begin{eqnarray}
&&\left(\begin{array}{lll}
\hat{A}_1\\ \hat{A}_2 \\ \hat{A}_3
\end{array}\right)= \left(\begin{array}{ccc}
\frac{1}{\sqrt{2}} & \frac{i}{\sqrt{2}} &0 \\
\frac{1}{\sqrt{2}} & -\frac{i}{\sqrt{2}} &0 \\
0&0&1
\end{array}\right)
\left(\begin{array}
{lll}
\hat{a}_1\\ \hat{a}_2 \\ \hat{a}_3
\end{array}\right)
\end{eqnarray}
and $\hat{\xi}_\alpha\equiv \hat{A}_\alpha \frac{1}{\sqrt{\hat{N}}}$.
It is easy to check that 
\begin{equation}
[\hat{A}_\alpha, \hat{A}_\beta^\dagger] = \delta_{\alpha\beta}. \quad \hat{A}_\alpha^\dagger\hat{A}_\alpha= \hat{a}_\alpha^\dagger\hat{a}_\alpha=\hat{N},\quad \hat{\xi}^\dagger_\alpha\hat{\xi}_\alpha=1.
\end{equation}
In terms of these new oscillators, 
\begin{eqnarray}
\hat{L}_+= \sqrt{2}(\hat{A}_3^\dagger  \hat{A}_1  - \hat{A}_2^\dagger \hat{A}_3), \quad \hat{L}_-= \hat{L}_+^\dagger, \quad \quad\mathrm{and}\quad
\hat{L}_3 =  (\hat{A}_1^\dagger\hat{A}_1 -\hat{A}_2^\dagger\hat{A}_2).
\end{eqnarray}
$\hat{L}_\pm, \hat{L}_3$ are defined in the appendix (\ref{appendix1}, \ref{appendix2}, \ref{appendix3}).
The operator $\hat{h}= N'_{\tilde{l}\tilde{n}}(\hat{\xi}_2^\dagger)^{\tilde{l}}(\hat{\xi}_1)^{\tilde{n}}$ satisfies 
\begin{eqnarray}
&&\hat{\mathcal{L}}_+ \hat{h} \equiv [\hat{L}_+ , \hat{h}]=0 \\
&& \hat{\mathcal{L}}_3 \hat{h} \equiv [\hat{L}_3 , \hat{h}]=\left(\tilde{l}+\tilde{n}\right) \hat{h}
\end{eqnarray} 
making $\hat{h}$ is the highest weight vector of the $SU(2)$  representation
with $j=(\tilde{l}+\tilde{n})$. 
 We denote this highest weight vector by $\Phi^j_{J,\kappa,j}$. The lower weight vectors can be obtained by the action of $ \hat{\mathcal{L}}_-$:
\begin{equation}
 (\hat{\mathcal{L}}_-)^{(j-m)} \Phi^j_{J,\kappa,j} = N_{J\kappa j m} \Phi^j_{J,\kappa,m}, \quad N_{J\kappa j m} =\mathrm{constant}.
\end{equation}
$\tilde{n}$ takes values $0,1\ldots n$. So  $j =(\tilde{l}+\tilde{n})$ takes all integer value from $\kappa$ to $J$:
\begin{equation}
j=\frac{\kappa}{2}, \frac{\kappa}{2}+2, \frac{\kappa}{2}+4\ldots J.
\end{equation}

Thus $\mathcal{H}_{nl}$ is spanned by the operators 
\begin{equation}
\Phi^j_{J,\kappa,m}  \quad\quad \mathrm{with}\quad  -j\leq m\leq j \quad j=\frac{\kappa}{2}, \frac{\kappa}{2}+1, \ldots  J.
\end{equation}
An arbitary element $\Phi$ of $\mathcal{H}_{nl}$ can be expressed as
\begin{equation}
\Phi= \sum_{j=\frac{\kappa}{2}}^J \sum_{m=-j}^j c^j_{J,\kappa,m}\Phi^j_{J,\kappa,m}, \quad c^j_{J,\kappa,m} \in \mathbb{C}.
\end{equation}

Any element $\Phi$ of $\mathcal{H}_{nl}$ is also an eigenfunction of the topological charge operator $\hat{K}_0$
\begin{equation}
\hat{K}_0\equiv[\hat{N},\quad], \quad\quad \hat{K}_0\Phi\equiv [\hat{N},\Phi]= \frac{\kappa}{2} \Phi.
\end{equation}
$\Phi$ is thus the noncommutative analogue of a section of the complex line bundle with topological charge $\kappa$, which takes only even integer values  ($\frac{\kappa}{2}\in \mathbb{Z}_+$).

\vspace{1in}

\appendix
\section{Appendix: Miscellaneous}
Using the map (\ref{map}), the condition (\ref{casimir22}) can be computed explicitly as
\begin{eqnarray}
y_i y_i &=& -\epsilon_{ijk}\epsilon_{ilm} \bar{z}_jz_k \bar{z}_l z_m \\
&=& -\left(\delta_{jl}\delta_{km}-\delta_{jm}\delta_{kl}\right)\bar{z}_jz_k \bar{z}_l z_m \\
&=& \bar{z}_jz_k \bar{z}_k z_j - \bar{z}_jz_k \bar{z}_j z_k \\
&=& \left(\bar{z}_j z_j\right)  (\bar{z}_k z_k)- (\bar{z}_j \bar{z}_j) (z_k z_k) \\
&=& (\bar{z}_j z_j)  (\bar{z}_k z_k)- \bar{\mathcal{O}} \mathcal{O}
\end{eqnarray}


The fuzzy computation is also straightforward.
Here we have to use the map (\ref{map2}). By direct substitution we get (\ref{casimir21}):
\begin{eqnarray}
\hat{C}=\hat{L}_i\hat{L}_i &=& -\epsilon_{ijk}\epsilon_{ilm} \hat{a}_j^\dagger \hat{a}_k \hat{a}_l^\dagger \hat{a}_m \\
&=&  -\left(\delta_{jl}\delta_{km}-\delta_{jm}\delta_{kl}\right)\hat{a}_j^\dagger \hat{a}_k \hat{a}_l^\dagger \hat{a}_m \\
&=& \hat{a}_j^\dagger \hat{a}_k \hat{a}_k^\dagger \hat{a}_j -\hat{a}_j^\dagger \hat{a}_k \hat{a}_j^\dagger \hat{a}_k \\
&=& \hat{a}_j^\dagger \hat{a}_k \left(\hat{a}_j \hat{a}_k^\dagger-\delta_{jk}\right) - \hat{a}_j^\dagger  \left(\hat{a}_j^\dagger\hat{a}_k + \delta_{jk}\right)\hat{a}_k \\
&=&  \left(\hat{a}_j^\dagger \hat{a}_j\right)\left(  \hat{a}_k \hat{a}_k^\dagger\right) -  2\hat{a}_j^\dagger \hat{a}_j- \left(\hat{a}_j^\dagger \hat{a}_j^\dagger\right)\left(\hat{a}_k\hat{a}_k\right) \\
&=& \hat{N}\left(\hat{N}+3\right) -2\hat{N} -\hat{\mathcal{O}}^\dagger\hat{\mathcal{O}}\\
&=& \hat{N}\left(\hat{N}+1\right)  -\hat{\mathcal{O}}^\dagger\hat{\mathcal{O}}
 \end{eqnarray}

The operators $\hat{A}_\alpha$ are defined as
\begin{eqnarray}\left(\begin{array}{lll}
\hat{A}_1\\ \hat{A}_2 \\ \hat{A}_3
\end{array}\right)= \left(\begin{array}{ccc}
\frac{1}{\sqrt{2}} & \frac{i}{\sqrt{2}} &0 \\
\frac{1}{\sqrt{2}} & -\frac{i}{\sqrt{2}} &0 \\
0&0&1
\end{array}\right)\left(\begin{array}
{lll}
\hat{a}_1\\ \hat{a}_2 \\ \hat{a}_3
\end{array}\right).
\end{eqnarray}

 It can be easily shown that these operators $\hat{A}_\alpha$ satisfy the oscillator algebra:
\begin{eqnarray}
&&[\hat{A}_1, \hat{A}_1^\dagger] = \frac{1}{2}([\hat{a}_1, \hat{a}_1^\dagger]+ [\hat{a}_2, \hat{a}_2^\dagger]) =1,\\
&&[\hat{A}_1, \hat{A}_2]=0, \\
&&[\hat{A}_1, \hat{A}_2^\dagger] = \frac{1}{2}([\hat{a}_1, \hat{a}_1^\dagger]- [\hat{a}_2, \hat{a}_2^\dagger]) =0.
\end{eqnarray}
The total number operator  can be reexpressed  in terms of the operators $\hat{A}_\alpha$ as
\begin{eqnarray} 
\hat{N} = \hat{a}_1^\dagger \hat{a}_1 + \hat{a}_2^\dagger \hat{a}_2+\hat{a}_3^\dagger \hat{a}_3 
= \hat{A}_1^\dagger \hat{A}_1 + \hat{A}_2^\dagger \hat{A}_2+\hat{A}_3^\dagger \hat{A}_3.
\end{eqnarray}
The operators $\hat{L}_i$ can also be rewritten in terms of the operators $\hat{A}_\alpha$ as
\begin{eqnarray}
\hat{L}_+&=&\hat{L}_1+i \hat{L}_2=\sqrt{2}(\hat{A}_3^\dagger  \hat{A}_1  - \hat{A}_2^\dagger \hat{A}_3),\label{appendix1} \\
\hat{L}_-&=&\hat{L}_1-i \hat{L}_2=\sqrt{2}(\hat{A}_1^\dagger  \hat{A}_3  - \hat{A}_3^\dagger \hat{A}_2),\label{appendix2}\\
\hat{L}_3& =& -(\hat{A}_1^\dagger\hat{A}_1 -\hat{A}_2^\dagger\hat{A}_2).\label{appendix3}
\end{eqnarray}

\end{document}